\documentclass[prd,showpacs,showkeys,amsmath,amssymb]{revtex4}
\usepackage{amsmath}
\usepackage{bm}
\usepackage{amssymb}
\usepackage{mathrsfs}
\usepackage[T1]{fontenc}
\usepackage{color}
\usepackage{braket}

\newcommand{\bea}{\begin{eqnarray*}}
\newcommand{\eea}{\end{eqnarray*}}
\newcommand{\bean}{\begin{eqnarray}}
\newcommand{\eean}{\end{eqnarray}}
\newcommand{\eqs}[1]{Eqs. (\ref{#1})}
\newcommand{\eq}[1]{Eq. (\ref{#1})}
\newcommand{\meq}[1]{(\ref{#1})}
\newcommand{\grad}{\nabla}

\newcommand{\non}{\nonumber \\}
\newcommand{\hsp}{\hspace{0.1mm}}
\newcommand{\pp}{\partial}
\newcommand{\sgt}{\sqrt{h}}

\begin{document}

\title{Consistency between dynamical and thermodynamical stabilities for charged self-gravitating perfect fluid}
\author{Wei Yang}
\author{Xiongjun Fang\footnote{Corresponding author: fangxj@hunnu.edu.cn}}
\author{Jiliang Jing}
\affiliation{Department of Physics, Key Laboratory of Low Dimensional Quantum Structures and
Quantum Control of Ministry of Education, and Synergetic Innovation Center for Quantum Effects and Applications, Hunan Normal
University, Changsha, Hunan 410081, P. R. China}

\begin{abstract}
\begin{center}
ABSTRACT
\end{center}
The entropy principle shows that, for self-gravitating perfect fluid, the Einstein field equations can be derived from the extrema of the total entropy, and the thermodynamical stability criterion are equivalent to the dynamical stability criterion. In this paper, we recast the dynamical criterion for the charged self-gravitating perfect fluid in Einstein-Maxwell theory, and further give the criterion of the star with barotropic condition. In order to obtain the thermodynamical stability criterion, first we get the general formula of the second variation of the total entropy for charged perfect fluid case, and then obtain the thermodynamical criterion for radial perturbation. We show that these two stability criterion are the same, which suggest that the inherent connection between gravity and thermodynamic even when the electric field is taken into account.
\end{abstract}


\maketitle

\section{Introduction}
It is well known that the black hole thermodynamics established the connection between gravity and thermodynamics. However, the origin of the black hole entropy is still the most important issue. In recent decades, a series of works have directly discussed the relationship between gravity and thermodynamics. For example, in 1994 Jacobson considered that the Einstein equation is a state equation on local Rindler causal horizons \cite{Jacobson1}. Then he also discussed how to construct Einstein equation from the equilibrium of total entanglement entropy in a "Causal Diamond" \cite{Jacobson2}. Both of these works beautifully give the in-depth relationship between Einstein equation and thermodynamics. In recent years, the proof of the maximum entropy principle have shown that, for self-gravitating perfect fluid, the gravitational equation can be derived from the constraint equation and the thermodynamic relation \cite{wald1981,gao,fanggr,fangEM,Cao1}. It has been proved that this principle not only applicable to general relativity, but also to other modified gravity theories, such as Lovelock gravity and $f(R)$ theories \cite{Cao2,fangfR}.

The relationship between gravity and thermodynamic is not only at the lowest order, i.e., the connection between the first order variation of the total entropy and the gravitational field equation. Considering their relationship in higher order is an interesting issue. Note that the dynamical stability criterion can be obtained by the first order variation of the gravitational equation. And the dynamical stable implies that all physical gauge invariant quantities remain bounded in time under linear perturbation of the star. Chandrasekhar first discussed dynamical stability by considering that all perturbed properties contain a factor $e^{i\omega t}$, and $\omega^2>0$ means that the star is dynamical stable \cite{Chandrasekhar}. Then Friedmann investigated that the "canonical energy" can provide the dynamical stability criterion in Lagrangian displacement framework \cite{Friedman4}, based on the previous works of Chandrasekhar, Friedman and Schutz \cite{Friedman1,Friedman2,Friedman3}. Seifert and Wald developed a general method to obtain the dynamical stability criterion for spherically symmetric perturbation in diffeomorphism covariant theories \cite{wald2007}. On the other hand, for thermodynamical stability of the star, the stability criterion is given by the second variation of total entropy is negative, $\delta^2S<0$, for all variation at fixed total physical properties. In fact, Cocke first pointed out that the thermodynamical stability criterion should take the same form as the corresponding dynamical stability criterion. Green et al. carefully studied the dynamical stability and thermodynamical stability of perfect fluid star, and presented that the necessary condition of thermodynamical stability is the positivity of canonical energy $\mathcal{E}$, which correspond the dynamical stability criterion \cite{wald2013}. Roupas proved that the maximum of total entropy of perfect fluid is consistent with the dynamical stability obtained by Yabushita \cite{Roupas,Yabu}. More recently, we show that the dynamical stability and thermodynamical stability are exactly the same both in general relativity or some modified theory such as $f(R)$ theories \cite{equgr,equfR}.

All the discussions of the relationship between thermodynamical and dynamical stability is still limited to the purely-gravitational theories. In order to investigate whether the entropy principle is valid for more general cases, we study the dynamical stability and thermodynamical stability for charged perfect fluid in Einstein-Maxwell theory. If the perfect fluid with no charge, considering that the entropy density $s$ is taken to be the function of energy density $\rho$ and particle number density $n$, and the ratio of the temperature $T$ to the chemical potential $\mu$ is constant, then we presented that the first and second variation of total entropy $S$ have a unified structure in different purely-gravitational theories \cite{equgr}. However, for charged perfect fluid in Einstein-Maxwell theory, since the Maxwell field yields an additional term in conservation equation, which shows that the ratio of the temperature and the chemical potential is related to the electric potential. In order to overcome this problem we introduced a new constant, and showed that the first variation of total entropy takes a different form compare with the uncharged case \cite{fangEM}. In this paper, taking the second variation of total entropy on the premise that the charged star is thermodynamic stable, through a naturally condition that the total particle number $N$ fixed, the constant we introduced would automatically eliminate. Then we obtain the specific structure of the thermodynamical stability criterion. Considering the star with the "barotropic" equation of the state, it is showed that the thermodynamical stability and dynamical stability criteria are actually equivalent. Our work suggest that the entropy principle remains valid not only for gravitational field but also with the electromagnetic field.

The rest of this paper is organized as follows. In the second section, we briefly recast the process of how to obtain the dynamical stability criterion for radial perturbation of the charged perfect fluid star. In the third section, together with the basic thermodynamic relations, we first obtain the thermodynamical stability criterion by the second variation of the equilibrium equation, then get the specific expression in the case of spherical symmetric perturbation, and show that these two stability criteria are exactly the same under the "barotropic" condition. Throughout our discussion, units will be used in which $c=G=1$, $\dot X$ and $X'$ are denoted as the derivative with respect to coordinate $t$ and $r$, respectively.

\section{Dynamical method}

In this section, we will briefly recast how to obtain the stability criterion for charged perfect fluid star by dynamical method. Our result is similar to although slightly different from Ref. \cite{Fernando}. To compare with the final result derived by thermodynamical method in next section, we would use the barotropic condition to simplify the show that our result seems more exact and reasonable. Considering a metric of spherical spacetime,
\begin{align}
\label{bgmetric}
ds^2=-e^{2\Psi(r,t)}dt^2+e^{2\Lambda(r,t)}dr^2+r^2d\theta^2+r^2\sin^2\theta d\psi^2 .
\end{align}
Total energy-momentum tensor of a charged perfect fluid takes the form
\begin{align}
\label{Tab}
T^{ab}=(\rho+p)u^{a}u^{b}+pg^{ab}
+\frac{1}{4\pi}\left(F^{ac}F^{b}\hsp_{c}-\frac{1}{4}g^{ab}F_{cd}F^{cd}\right) ,
\end{align}
and the Maxwell's equation with source reads as
\begin{align}
\label{Maxwell}
\nabla_{b}F^{ab}=4\pi j^{a}, \quad \nabla_{[a}F_{bc]}=0 ,
\end{align}
where $j^a$ is the 4-current density. For spherically symmetric case the only non vanishing component of the Maxwell's tensor is \cite{waldbook}
\bean
\label{Fab}
F^{01}=e^{-\Psi-\Lambda}\frac{Q(r,t)}{r^2} ,
\eean
where $Q(r,t)=\int^{r}_{0}e^{\Psi+\Lambda}4\pi r^2j^{0}dr$ is the total electric charge inside the sphere of radius $r$ at time $t$. The static background give the components of Einstein equation
\begin{align}
\label{G00}
8\pi T^{0}\hsp_{0}&=\frac{Q^2}{r^4}+8\pi\rho=e^{-2\Lambda}\left(\frac{2\Lambda'}{r}-\frac{1}{r^2}\right)+\frac{1}{r^2} , \\
\label{G11}
8\pi T^{1}\hsp_{1}&=\frac{Q^2}{r^4}-8\pi p=-e^{-2\Lambda}\left(\frac{2\Psi'}{r}+\frac{1}{r^2}\right)+\frac{1}{r^2} ,
\end{align}
and the $r$ component of conservation equation $\grad_aT^a\hsp_b=0$ gives
\bean
\label{drpsi}
p'=\frac{Q Q'}{4\pi r^4}-(p+\rho)\Psi' .
\eean

Now we consider a small radial perturbation. The Eulerian perturbation of any tensor quantities $\delta X$ and the Lagrangian perturbation $\Delta X$ satisfied the relation \cite{wald2014}
\bean
\label{perX}
\Delta{X}=\delta X+\mathcal{L}_\xi X ,
\eean
where $\xi$ is the Lagrangian displacement. Since $u^r/u^\tau=\dot\xi$, the four velocity in the perturbed configuration are given by \cite{Chandrasekhar}
\bean
u^0=e^{-\Psi}(1-\delta\Psi), \quad u^1=e^{-\Psi}\dot\xi
\eean
Assuming that the charge distribution remains unchanged in the oscillating configuration, $\Delta Q(r,t)=0$, which implies that there are no electric currents for comoving observer. This assumption gives $\delta Q=-\xi\cdot Q'$. Together with this relation, the $rr$ component of the perturbed Einstein equation gives
\bean
\label{delG11}
\frac{2Q Q'\xi}{r^4}+8\pi\delta p
=-e^{-2\Lambda}\left(\frac{4}{r}\Psi'+\frac{2}{r^2}\right)\delta\Lambda+e^{-2\Lambda}\frac{2}{r}\delta\Psi' ,
\eean
and the integrated of $tr$ component of the Einstein equation gives
\bean
\label{delLambda}
\delta\Lambda=-4\pi re^{2\Lambda}\xi\left(p+\rho\right)=-(\Lambda'+\Psi')\xi .
\eean
Together with \eq{delLambda}, the baryon conservation $\grad_a(nu^a)=0$ yields
\bean
\label{deln}
\delta n = -\frac{e^{\Psi}}{r^2} \frac{\pp}{\pp{r}}\left(n r^2\xi e^{-\Psi}\right), \quad\quad
\Delta n = -\frac{e^{\Psi}}{r^2} n \frac{\pp}{\pp{r}}\left(r^2\xi e^{-\Psi}\right).
\eean
The adiabatic condition $\Delta p/\Delta n=\gamma\cdot p/n$ \cite{Chandrasekhar}, where $\gamma$ is the ratio of the specific heat, yields
\bean
\label{delp}
\delta{p}=\Delta p+\xi p'=-\gamma p\frac{e^{\Psi}}{r^2}\frac{\pp}{\pp{r}}\left(r^2\xi e^{-\Psi}\right)-\xi p' .
\eean
Substituting \eqs{delLambda} and \meq{delp} into \eq{delG11} to eliminate $\delta\Lambda$ and $\delta p$, we obtain
\bean
\label{delpsir}
\delta\Psi'
=-e^{2\Lambda_0}\frac{Q Q'\xi}{r^3}-4\pi\gamma p\frac{e^{2\Lambda+\Psi}}{r}\left(r^2\xi e^{-\Psi}\right)'
+4\pi e^{2\Lambda}\xi [r p'-(p+\rho)] .
\eean

Then we consider the perturbed conservation equation. Projecting this equation parallelled to the matter flow, i.e. $u^a \grad_b T^b\hsp_a=0$, gives
\bean
\label{Proflow}
-\frac{d\rho}{d\tau}+\frac{\left(p+\rho\right)}{{n}}\frac{d{{n}}}{d\tau}-\frac{Q}{4\pi r^4}\frac{dQ}{d\tau}=0 ,
\eean
which yields
\begin{align}
\label{dyndelrho}
\delta{\rho}=-(\rho+p)\frac{e^{\Psi}}{r^2}\left(r^2\xi e^{-\Psi}\right)'-\xi\rho' .
\end{align}
While the $r$-component of projecting transversely to the matter flow, $\left(\frac{\pp}{\pp r}\right)^a(\delta^b\hsp_a+u^bu_a)\grad_c T^c\hsp_b=0$, gives the relativistic Euler equation
\bean
\label{RelEuler}
\left(p+\rho\right)e^{2\Lambda-2\Psi}\ddot{\xi}=-\left(\delta{p}+\delta{\rho}\right)\Psi'
-\left(p+\rho\right)\delta\Psi'-(\delta p)'-\frac{Q Q'\xi'}{4\pi r^4}-\frac{(Q')^2\xi}{4\pi r^4}-\frac{Q Q''\xi}{4\pi r^4 } .
\eean
It should be noted that comparing with Eq. (27) of Ref. \cite{Fernando}, we have an additional part which is the last term of the righthand of the above equation.

Substituting \eqs{delp}, \meq{delpsir} and \meq{dyndelrho} into \eq{RelEuler} to rewrite all perturbed properties in terms of $\xi$, we obtain
\begin{equation}
\begin{split}
\label{ddotxi}
(p+\rho)e^{2\Lambda-2\Phi}\ddot{\xi} =& \left[\gamma p\frac{e^{\Psi}}{r^2}\left(r^2e^{-\Psi}\xi\right)'+\xi p'\right]'
+\left[(p+\gamma p+\rho) \frac{e^{\Psi}}{r^2}\left(r^2 e^{-\Psi}\xi\right)'+(p'+\rho')\xi \right]\Psi' \\
&\left.+(p+\rho)\left[e^{2\Lambda}\frac{Q Q'\xi}{r^3}+4\pi\gamma p\frac{e^{2\Lambda+\Psi}}{r}\left(r^2 e^{\Psi}\xi\right)'
+4\pi e^{2\Lambda} (p+\rho-rp')\xi\right]\right.\\
& -\frac{Q Q'}{4\pi r^4}\xi'-\frac{(Q')^2}{4\pi r^4}\xi-\frac{Q Q''\xi}{4\pi{r^4}} .
\end{split}
\end{equation}
Assuming that $\xi(r,t)=\xi(r)e^{-i\omega t}$, the \eq{ddotxi} becomes the pulsation equation. Multiplying $r^2 e^{\Lambda+\Phi}\xi$ and integrating the equation over the range $r$ from $0$ to the radius of the star $R$, considering the system satisfied the boundary condition $\xi=0$ at $r=0$ and $\delta p=0$ at $r=R$ \cite{Chandrasekhar}, after some directly calculation it can be obtained that
\bean
\label{dyn}
\omega^2 \int_0^R e^{3\Lambda-\Phi}(p+\rho)\xi^2 = L ,
\eean
where
\begin{equation}
\begin{split}
\label{L}
L=& \int_0^r e^{3\Psi+\Lambda}\gamma p r^{-2}\left[\frac{d}{dr}(r^2e^{-\Psi}\xi)\right]^2
+4re^{\Psi+\Lambda}\frac{dp}{dr}\xi^2 +8\pi r^2 p (p+\rho)e^{\Psi+3\Lambda}\xi^2 \\
& -e^{\Psi+\Lambda}r^2\left(\frac{d\Psi}{dr}\right)^2\left(p+\rho\right)\xi^2-e^{\Psi+3\Lambda}\frac{Q^2}{r^2}\left(p+\rho\right)\xi^2 .
\end{split}
\end{equation}
\eq{dyn} is the dynamical stability for charged perfect fluid. This result would degenerate to the result of Chandrasekhar's \cite{Chandrasekhar} if $Q(r)$ vanishes. Note that in Ref. \cite{Fernando} it has defined the "renormalized displacement function" $\zeta=r^2e^{-\Phi}\xi$ to reexpress this result. However, in what issue we are considering is that to compare the dynamical stability and thermodynamical stability, there is no need to define the new variable to rewritten \eq{dyn}. Since Roupas pointed out that for uncharged perfect fluid, the dynamical stability would equivalent to thermodynamical stability only under the barotropic condition \cite{Roupas}. We consider that all particles possess the same charge, and the Lagrangian for the perfect fluid takes the form
\bean
\label{Lmat}
\mathcal{L}_{mat}=-\varrho(n) ,
\eean
where $n$ is the particle number density. It is showed that there exists an identification \cite{wald2007}
\bean
\label{varrho}
\rho\rightarrow \varrho, \quad  p\rightarrow\frac{\pp{\varrho}}{\pp{n}}n-\varrho,
\eean
which yields
\bean
\label{delvarrho}
\delta{\rho}=\frac{\pp{\varrho}}{\pp{n}}\delta{n}, \quad  \delta{p}=\frac{\pp^2\varrho}{\pp{n}^2}n\delta{n},
\eean
and
\bean
\gamma p=(p+\rho)\frac{\pp p}{\pp n}\left(\frac{\pp\rho}{\pp n}\right)^{-1}=\frac{\pp^2\varrho}{\pp n^2} n^2.
\eean

Together with these relations and substituting the $\theta\theta$ component of field equation
\bean
\frac{Q^2}{r^4}+8\pi{p}=e^{-2\Lambda}\left[\Psi''-\Psi'\Lambda'+(\Psi')^2+\frac{1}{r}(\Psi'-\Lambda')\right] ,
\eean
into \eq{dyn}, the dynamical stability criterion takes the form
\begin{equation}
\label{Dynfinal}
\omega^2\int_0^re^{3\Lambda-\Psi}(p+\rho)r^2\xi^2 dr = \int_0^r \hat L dr ,
\end{equation}
where
\begin{equation}
\begin{split}
\hat L =& \frac{e^{3\Psi+\Lambda}}{r^2}\frac{\pp^2\varrho}{\pp{n}^2}n\left[\frac{d}{dr}(r^2e^{-\Psi}\xi)\right]^2+\frac{4e^{\Psi+2\Lambda}}{r} n qQ \xi^2
 -(\varrho'n)e^{\Psi+\Lambda}r^2\left(\Psi'\Lambda'-\Psi''+\frac{3\Psi'}{r}+\frac{\Lambda'}{r}\right)\xi^2 .
\end{split}
\end{equation}
For the left hand of \eq{Dynfinal}, the coefficient of $\omega^2$ are actually the inner product and positive definite \cite{wald2007}. Hence if the signature of $\int_0^R\hat L dr$ is positive, then the star is dynamical stable.

\section{Thermodynamic method}
In this section, we would present the expression of the second variation of total entropy for charged perfect fluid. Then considering the radial perturbation to obtain the thermodynamical stability criterion. For self-gravitating perfect fluid over any selected region $C$ on $\Sigma$, consider that the ordinary thermodynamic relation and the Tolman's law $T\chi=const.$ are always satisfied, where $T$ and $\chi$ are the temperature of the fluid and the redshift factor of the static observer. It is showed that if the Lagrangian constructed purely by the metric and its derivatives, the second variation of total entropy under the perturbation can be written as a uniform expression \cite{equgr,equfR},
\bean
\label{del2Snocharge}
\delta^2S=\int_C\frac{1}{T}\left[2\delta\rho\delta\sgt+\sgt\delta^2\rho+(p+\rho)\delta^2\sgt-\frac{\delta p \delta\rho}{p+\rho}\sgt \right] ,
\eean
where $T$ is the temperature of the fluid and $h$ is the determinant of the induced metric $h_{ab}$ on $\Sigma$. This relation is derived from the first variation of total entropy for uncharged star. However, if the perfect fluid with charge, the expression of first variation of total entropy should be modified. The local first law of thermodynamics and the Gibbs-Duhem relation give
\bean
\label{therfirst}
T ds= d\rho-\mu dn, \quad  p+\rho=\mu n+ Ts ,
\eean
where $s$ and $\mu$ are the entropy density and the chemical potential, respectively. Assuming that all particle possess the same charge $q$, the conservation equation of charged perfect fluid yields \cite{fangEM}
\bean
\label{muT}
\frac{\mu}{T}+q\Phi=\beta ,
\eean
where $\beta$ is a constant, $\Phi$ is the electrostatic potential defined as $\Phi=-A_a\tilde\xi^a$, $A_a$ and $\tilde\xi^a$ are the vector potential and killing vector, respectively. Without loss of generality, consider that $T\chi=1$. Together with \eqs{therfirst} and \meq{muT}, we get
\bean
dp=\left(\frac{p+\rho}{T}\right)dT-n q T d\Phi .
\eean

Applying these relation, it is showed that the first variation of total entropy takes the form \cite{fangEM}
\bean
\label{delS}
\delta{S}=\int_C\frac{1}{T}\left(p+\rho-(\beta-q\Phi)Tn\right)\delta\sqrt{h}+\sqrt{h}\left(q\Phi-\beta\right)\delta{n}+\sqrt{h}\frac{1}{T}\delta\rho ,
\eean
and it can be proved that \eq{delS} is equivalent to the Einstein-Maxwell field equation. Since $\delta S=0$ implies that the system is in thermal equilibrium, to investigate whether this equilibrium is a stable equilibrium, we should calculate the second variation of total entropy $\delta^2S$. And if $\delta^2S<0$, we can say that the isolated star is thermodynamical stable. Taking the variation on \eq{delS} and considering that the Tolman's law remains valid, we obtain the second variation of total entropy takes the form
\begin{equation}
\begin{split}
\label{del2S}
\delta^2S=&\int_C\frac{2}{T}\delta{\rho}\cdot\delta\sqrt{h}+2\left(q\Phi-\beta\right)\delta n \cdot \delta\sqrt{h}+\sqrt{h}q\delta\Phi \cdot\delta{n} \\
& -\frac{\sqrt{h}\delta p \cdot \delta\rho}{T(p+\rho)}-\frac{\rho_e\sgt\delta\Phi \cdot \delta\rho}{p+\rho}
+\frac{1}{T}\left[p+\rho+(-\beta+q\Phi)Tn\right]\delta^2\sqrt{h} \\
& +\sqrt{h}(q\Phi-\beta)\delta^2{n}+\frac{\sqrt{h}}{T}\delta^2\rho .
\end{split}
\end{equation}
Note that in the above expression, there is a constant $\beta$ which can be chosen arbitrarily. However, the total number of particle $N$, which takes the integral form as $N=\int_C n\sgt$, can be considered that be fixed at any order. Taking the second variation of total particle number vanishes,
\bean
\label{del2N}
\delta^2{N}=\int_C\sqrt{h}\delta^2{n}+2\delta\sqrt{h}\delta{n}+{n}\delta^2\sqrt{h}=0 ,
\eean
and substituting this result into \eq{del2S}, we get a expression in which the constant $\beta$ is eliminated,
\begin{equation}
\begin{split}
\label{del2Sgen}
\delta^2S=&\int_C\frac{2}{T}\delta{\rho}\cdot\delta\sqrt{h}+2q\Phi\delta{n}\cdot\delta\sqrt{h}
+\sqrt{h}q\delta\Phi\cdot\delta{n}-\frac{\sqrt{h}\delta p \cdot\delta\rho}{T(p+\rho)}
-\frac{\sqrt{h}nq\delta\Phi\cdot\delta\rho}{p+\rho} \\
& +\frac{1}{T}\left(p+\rho+q\Phi Tn\right)\delta^2\sqrt{h}+\sqrt{h}q\Phi\delta^2{n}+\frac{\sqrt{h}}{T}\delta^2\rho .
\end{split}
\end{equation}
Using \eq{del2Sgen} one can directly obtain the specific form of the stability criterion.

To compare with the dynamical stability criterion under radial perturbation, now we consider the spherically symmetric perturbations of a charged fluid star. The relation $j^t=nq e^{-\Psi}$ yields $Q'=4\pi r^2nqe^{\Lambda}$, together with the assumption $\Delta Q=0$, hence $\delta Q=-Q'\xi=-4\pi r^2nqe^{\Lambda} \xi$. Using the relation \eq{delLambda}, the first variation of constraint equation, $\delta{G_{00}} = 8{\pi}\delta{T}_{00}$, gives
\bean
\label{delrho}
\delta\rho=-\frac{1}{r^2}\frac{\pp}{\pp{r}}\left[r^2\left(p+\rho\right)\xi\right]-\frac{2Q\delta{Q}}{8\pi r^4}
=-\frac{1}{r^2}\frac{\pp}{\pp{r}}\left[r^2\xi\left(p+\rho\right)\right]+\frac{e^{\Lambda}nqQ\xi}{r^2} .
\eean
With \eqs{delLambda} and \meq{deln}, the second variation of $\rho$ could be written as
\begin{equation}
\begin{split}
\label{del2rho}
\delta^2\rho=&-\frac{1}{r^2}\frac{\pp}{\pp{r}}\left[r^2\left(\delta{p}+\delta\rho\right)\xi+r^2(p+\rho)\delta\xi\right]+\frac{1}{r^2}e^{\Lambda}nqQ\delta\xi \\
& -4\pi e^{2\Lambda}n^2q^2\xi^2-\frac{4\pi}{r}nqQ\xi^2e^{3\Lambda}(p+\rho)-\frac{1}{r^4}e^{\Psi+\Lambda}qQ\xi\frac{\pp}{\pp{r}}(r^2e^{-\Psi}{n\xi}) .
\end{split}
\end{equation}
While the first and second variation of the induced metric could be written as
\bean
\label{delh}
\delta\sqrt{h}=e^{\Lambda}r^2\sin\theta\delta{\Lambda}=-4\pi e^{3\Lambda} r^3(p+\rho)\sin\theta\xi ,
\eean
and
\begin{equation}
\begin{split}
\label{del2h}
\delta^2\sqrt{h} =& 48\pi^2r^4e^{5\Lambda}(p+\rho)^2\sin\theta\xi^2
-4\pi e^{3\Lambda}r^3(\delta{p}+\delta{\rho})\sin\theta\xi \\
& -4\pi e^{3\Lambda}r^3(p+\rho)\sin\theta\delta\xi .
\end{split}
\end{equation}
According to \eq{deln}, the second order variational of $n$ can directly be given by
\bean
\label{del2n}
\delta^2{n}=n\xi{\delta\Psi'}+\frac{e^{\Psi}}{r^2}\xi'\frac{\pp}{\pp{r}}\left(r^2e^{-\Psi}n\xi\right)
+\frac{e^{\Psi}}{r^2}\xi\frac{\pp^2}{\pp{r}^2}\left(r^2e^{-\Psi}n\xi\right)-\frac{ e^{\Psi}}{r^2}\frac{\pp}{\pp{r}}\left(r^2e^{-\Psi}n\delta\xi\right) . \non
\eean
For spherical symmetry case, the integration $\int_C$ becomes as $4\pi\int_r dr$. Using integration by parts and dropping the boundary terms, we can get the explicit expression of each term of $\delta^2S$, which are presented in Appendix A. Together with \eqs{P1}$\sim$\meq{P8}, we find that the coefficient of $\delta\xi$ would vanish, and the second variation of total entropy can be written as
\begin{equation}
\begin{split}
\label{del2Sres}
\delta^2{S}=&4\pi\int_r4\pi{r^4}e^{\Psi+3\Lambda}(p+\rho)^2\left(\frac{2\Psi'}{r}+\frac{1}{r^2}\right)\xi^2-8\pi{re^{\Psi+4\Lambda}}n q Q (p+\rho)\xi^2\\
&+4\pi{r^3}q\Phi n'e^{3\Lambda}\left(p+\rho\right)\xi^2-4\pi{r^3}nq\Phi e^{3\Lambda}\left(p+\rho\right)'\xi^2
+4\pi{r}n^2q^2\Phi{Q} e^{4\Lambda} \xi^2 \\
& -4\pi{r^3}e^{3\Lambda}\left(p+\rho\right)nq\Phi\Psi'\xi^2-{r^2}n'qe^{\Lambda}\delta\Phi\cdot \xi+{r^2}nqe^{\Lambda}\Psi'\delta\Phi\cdot\xi \\
& +\frac{r^2nqe^{\Lambda}}{p+\rho}(p+\rho)'\delta{\Phi}\cdot\xi-\frac{e^{2\Lambda}n^2q^2Q}{p+\rho}\delta{\Phi}\cdot\xi
-r^2e^{\Psi+\Lambda}\frac{\delta{p}\delta\rho}{p+\rho}\\
&-4\pi{r^3}e^{3\Lambda}nq\Phi\xi\cdot\delta\rho-4{\pi}e^{\Psi+3\Lambda}r^2n^2q^2\xi^2\\
& -4\pi{r}q\Phi e^{\Psi+3\Lambda}(p+\rho)\xi\frac{\pp}{\pp{r}}\left(r^2e^{-\Psi}n\xi\right) .
\end{split}
\end{equation}
In the simplified calculation of $\delta^2S$, we find that the coefficient of $\delta\xi$ would vanishes, which is natural since we consider that the state under a perturbation deviates only slightly from the equilibrium state. Now using the identification \eq{delvarrho}, we get the relation
\bean
\frac{\pp n}{\pp r}=-\frac{\pp\varrho/\pp n}{\pp^2\varrho/\pp n^2}\Psi'+\frac{qQe^{\Lambda}}{r^2}\frac{1}{\pp^2\varrho/\pp n^2} ,
\eean
then a direct but tedious computations shows that the coefficient of $\delta\Phi$ would vanish, and the final expression of $\delta^2S$ can be given by
\begin{equation}
\begin{split}
\label{del2Sfinal}
\delta^2S
 =& 4\pi\int_0^R \Bigg\{ r^2e^{\Psi+\Lambda}(p+\rho)\xi^2\left(\frac{\pp\Psi}{\pp{r}}\frac{\pp{\Lambda}}{\pp{r}}
-\frac{d^2\Psi}{dr^2}+3\frac{\Psi'}{r}+\frac{\Lambda'}{r}\right)  \\
& -\frac{e^{3\Psi+\Lambda}}{r^2}\frac{\pp^2\varrho}{\pp{n}^2}n^2\left[\frac{\pp}{\pp{r}}\left(r^2e^{-\Psi}\xi\right)\right]^2
-e^{\Psi+2\Lambda}nqQ\xi^2\frac{4}{r} \Bigg\} dr .
\end{split}
\end{equation}
Hence we obtain the thermodynamical stability criterion for spherical symmetry perturbation for charged perfect fluid star.

\section{Conclusions and discussions}

It is well known that the thermodynamical stable for isolated star requires that the second variation of total entropy to be negative, $\delta^2S<0$. However, the star is in dynamical stable requires that the left hand of \eq{Dynfinal} to be positive. Comparing \eqs{Dynfinal} and \meq{del2Sfinal}, it is shown that the criterion for the dynamical stability is consistent with the dynamical stability criterion for radial perturbation of charged perfect fluid star with barotropic equation of state.

The proof of maximum entropy principle provide a solid connection between thermodynamics and gravity. The significant improvement from precious works is that what we discussed in this manuscript is no longer the situation with no charge. Our work suggests that for the theories in terms of not only the metric and its derivatives, the maximum entropy principle may still work.

\section*{Acknowledgments}
We thank Xiaokai He for some useful discussion. This work was supported by National Natural Science Foundation of China (NSFC) with Grants No. 11705053 and No. 12035005.

\appendix
\section{Explicit expression for each term of $\delta^2S$}
\label{appA}
In this appendix, we would give the explicit expression of $\delta^2S$ in spherical symmetric case. Note that in the following calculation, we use the integration by parts and drop the boundary terms, and also consider that the Tolman's law is also valid, $T^{-1}=\chi=e^{\Psi}$. We denote that the $n$-th term of the right hand of \eq{del2Sgen} as $P_n$, and simply write the integral $\int_0^R dr$ as $\int_r$. Together with \eqs{delrho} and \meq{delh}, we obtain
\begin{equation}
\begin{split}
\label{P1}
P_1 &= \int_C\frac{2}{T}\delta\rho\delta\sqrt{h} \\
&=4\pi\int_r 4\pi \frac{1}{r}e^{\Psi+3\Lambda}\frac{\pp}{\pp{r}}\left[r^2\xi\left(p+\rho\right)\right]^2
-8\pi{re^{\Psi+4\Lambda}}nqQ\xi^2(p+\rho)  \\
&=4\pi\int_r -48\pi^2{r^4}e^{\Psi+5\Lambda}(p+\rho)^3\xi^2+4\pi{r^4}e^{\Psi+3\Lambda}(p+\rho)^2\xi^2\left(\frac{2\Psi'}{r}+\frac{1}{r^2}\right)
-8\pi{re^{\Psi+4\Lambda}}nqQ\xi^2(p+\rho).
\end{split}
\end{equation}

Considering \eq{deln} and the relation $\Phi'=-e^{\Psi+\Lambda}\frac{Q}{r^2}$, we get
\begin{equation}
\begin{split}
\label{P2}
P_2=&\int_C2q\Phi\delta{n}\delta\sqrt{h}\\
=&4\pi\int_r4\pi\frac{q\Phi}{nr}e^{2\Psi+3\Lambda}\left(p+\rho\right)\frac{\pp}{\pp{r}}\left(r^2e^{-\Psi}n\xi\right)^2\\
=&4\pi\int_r4\pi{r}nqQe^{\Psi+4\Lambda}\xi^2\left(p+\rho\right)+4\pi{r^3}q\Phi\xi^2n'e^{3\Lambda}\left(p+\rho\right)
+4\pi{r^2}nq\Phi\xi^2e^{3\Lambda}\left(p+\rho\right) \\
& -4\pi{r^3}nq\Phi\xi^2e^{3\Lambda}\left(2\Psi'+3\Lambda'\right)\left(p+\rho\right)-4\pi{r^3}nq\Phi\xi^2e^{3\Lambda}\left(p+\rho\right)' ,
\end{split}
\end{equation}
and
\begin{equation}
\begin{split}
\label{P3}
P_3 &= \int_C\sqrt{h}q\delta\Phi\delta{n} \\
&=4\pi\int_r-qe^{\Psi+\Lambda}\delta\Phi\frac{\pp}{\pp{r}}\left(r^2e^{-\Psi}n\xi\right)\\
&=4\pi\int_r-2rnqe^{\Lambda}\delta{\Phi}\xi-{r^2}n'qe^{\Lambda}\delta\Phi\xi+{r^2}nqe^{\Lambda}\delta\Phi\xi\Psi'-{r^2}nqe^{\Lambda}\delta\Phi\xi' .
\end{split}
\end{equation}

The fourth term $P_4$ and the fifth term $P_5$ can easily be showed as
\begin{equation}
\begin{split}
\label{P4}
P_4 = \int_C-\frac{\sqrt{h}\delta{p}\delta\rho}{T(p+\rho)}
=4\pi\int_r-r^2e^{\Psi+\Lambda}\frac{\delta{p}\delta\rho}{p+\rho} .
\end{split}
\end{equation}
\begin{equation}
\begin{split}
\label{P5}
P_5 &= \int_C-\frac{\sqrt{h} n q \delta\Phi\delta\rho}{p+\rho} \\
&=4\pi\int_r2rnqe^{\Lambda}\delta{\Phi}\xi+r^2nqe^{\Lambda}\delta{\Phi}\xi'+\frac{r^2nqe^{\Lambda}\delta{\Phi}}{p+\rho}\xi\left(p+\rho\right)'
-\frac{e^{2\Lambda}n^2q^2Q\xi\delta{\Phi}}{p+\rho} .
\end{split}
\end{equation}

Using \eqs{del2h}, we have
\begin{equation}
\begin{split}
\label{P6}
P_6 =&\int_C\frac{1}{T}\left(p+\rho+q\Phi{Tn}\right)\delta^2\sqrt{h}\\
=&4\pi\int_r 48\pi^2r^4e^{\Psi+5\Lambda}\xi^2\left(p+\rho\right)^3-4\pi{r^3}e^{\Psi+3\Lambda}\xi\left(p+\rho\right)\left(\delta{p}+\delta\rho\right)
-4\pi{r^3}e^{\Psi+3\Lambda}\left(p+\rho\right)^2\delta\xi\\
& +48\pi^2r^4e^{5\Lambda}nq\Phi\xi^2\left(p+\rho\right)^2 -4\pi{r^3}e^{3\Lambda}nq\Phi\xi\left(\delta{p}+\delta\rho\right)
-4\pi{r^3}e^{3\Lambda}\left(p+\rho\right)nq\Phi\delta\xi .
\end{split}
\end{equation}

Together with \eqs{delpsir} and \meq{del2n}, we get
\begin{equation}
\begin{split}
\label{P7}
P_7 =& \int_C\sqrt{h}q\Phi\delta^2{n} \\
=& 4\pi\int_r4\pi{r}n^2q^2\Phi{Q}\xi^2{e^{4\Lambda}}+4\pi{r^3}nq{\Phi}e^{3\Lambda}\delta p \xi
-4\pi{r^4}e^{3\Lambda}\xi^2\left(p+\rho\right)nq\Phi\left(\frac{2\Psi'}{r}
+\frac{1}{r^2}\right) \\
&+q\Phi\xi'e^{\Psi+\Lambda}\frac{\pp}{\pp{r}}\left(r^2e^{-\Psi}n\xi\right)+q\Phi{\xi}e^{\Psi+\Lambda}\frac{\pp^2}{\pp{r^2}}\left(r^2e^{-\Psi}n\xi\right)
-{q\Phi}e^{\Psi+\Lambda}\frac{\pp}{\pp{r}}\left(r^2e^{-\Psi}n\delta\xi\right) .
\end{split}
\end{equation}

And \eq{del2rho} yields
\begin{equation}
\begin{split}
\label{P8}
P_8 =&\int_C\frac{\delta^2\rho}{T}\sqrt{h}\\
=&4\pi\int_r4{\pi}e^{\Psi+3\Lambda}r^3\xi\left(\delta{p}+\delta\rho\right)(p+\rho)+4{\pi}e^{\Psi+3\Lambda}r^3\delta\xi(p+\rho)^2
+e^{\Psi+2\Lambda}nqQ\delta\xi\\
&-4{\pi}e^{\Psi+3\Lambda}r^2n^2q^2\xi^2-4\pi{r}nqQ\xi^2e^{\Psi+4\Lambda}(p+\rho)-\frac{1}{r^2}e^{2\Phi+2\Lambda}qQ\xi\frac{\pp}{\pp{r}}(r^2e^{-\Psi}{n\xi}) .
\end{split}
\end{equation}

\end{document}